\documentclass[aps,pra,reprint,superscriptaddress]{revtex4-2}
\usepackage{graphicx}
\usepackage{dcolumn}
\usepackage{bm}
\usepackage[utf8]{inputenc}
\usepackage[T1]{fontenc}
\usepackage{booktabs, array, mathptmx, float, tabularx, booktabs, lipsum, multirow, amstext}
\usepackage{amsmath}
\usepackage{mathptmx}
\usepackage{mathrsfs}
\DeclareMathAlphabet{\mathcal}{OMS}{cmsy}{m}{n}
\usepackage{siunitx, xcolor}
\usepackage[version=4]{mhchem}

\begin{document}
	
	\title{Pound-Drever-Hall Feedforward: Laser Phase Noise Suppression beyond Feedback}
	
	\author{Yu-Xin Chao}
	\thanks{These authors contributed equally to this work.}
	\affiliation{State Key Laboratory of Low Dimensional Quantum Physics, Department of Physics, Tsinghua University, Beijing 100084, China.}
	
	\author{Zhen-Xing Hua}
	\thanks{These authors contributed equally to this work.}
	\affiliation{State Key Laboratory of Low Dimensional Quantum Physics, Department of Physics, Tsinghua University, Beijing 100084, China.}
	
	\author{Xin-Hui Liang}
	\affiliation{State Key Laboratory of Low Dimensional Quantum Physics, Department of Physics, Tsinghua University, Beijing 100084, China.}
	
	\author{Zong-Pei Yue}
	\affiliation{State Key Laboratory of Low Dimensional Quantum Physics, Department of Physics, Tsinghua University, Beijing 100084, China.}
	
	\author{Li You}
	\affiliation{State Key Laboratory of Low Dimensional Quantum Physics, Department of Physics, Tsinghua University, Beijing 100084, China.}
	\affiliation{Frontier Science Center for Quantum Information, Beijing, China.}
	\affiliation{Collaborative Innovation Center of Quantum Matter, Beijing, China.}
	\affiliation{Hefei National Laboratory, Hefei, Anhui 230088, China.}
	
	\author{Meng Khoon Tey}
	\email{mengkhoon\_tey@mail.tsinghua.edu.cn}
	\affiliation{State Key Laboratory of Low Dimensional Quantum Physics, Department of Physics, Tsinghua University, Beijing 100084, China.}
	\affiliation{Frontier Science Center for Quantum Information, Beijing, China.}
	\affiliation{Collaborative Innovation Center of Quantum Matter, Beijing, China.}
	\affiliation{Hefei National Laboratory, Hefei, Anhui 230088, China.}
	\date{\today}
	
	\begin{abstract}
		Pound-Drever-Hall (PDH) laser frequency stabilization is a powerful technique widely used for building narrow-linewidth lasers. This technique is however ineffective in suppressing high-frequency (>100~kHz) laser phase noise detrimental for many applications. Here, we introduce an effective method which can greatly enhance its high-frequency performance. The idea is to recycle the residual PDH signal of a laser locked to a cavity, by feedforwarding it directly to the laser output field after a delay fiber. Using this straightforward method, we demonstrate a phase noise suppression capability about 4 orders of magnitude better than just using usual PDH feedback for phase noise around a few MHz. We further find that this method exhibits noise suppression performance equivalent to cavity filtering. The new method holds great promise for applications demanding highly stable lasers with diminished phase noise up to tens of MHz, e.g. precise and high-speed control of atomic and molecular quantum states.
	\end{abstract}
	
	\maketitle
	
	{\it Introduction. ---} 
	The emergence of narrow linewidth lasers frequency-locked to ultra-stable optical reference cavities has paved the way for a range of revolutionary technologies, including gravitational wave sensing~\cite{2015_LIGO_AdvancedLIGO}, optical clocks~\cite{2015_RMP_OpticalAtomicClock,2021Katori_portableClock,2019_YeJun_SrClock,2021_LinYiGe_SrClock,2019Leibrandt_PRL_Al}, ultra-low noise photonic microwave generation~\cite{2011Fortier_Optical_microwave,2017XieMicrowave,2020Quinlan_Science_Microwave,2020Kippenberg_photonicMicrowave}, high fidelity control of atomic qubits~\cite{2018_Browaeys_Coherence,2019_Levine_MultiqubitGate,2020_ManuelEndres_AKEA,2019_APR_Ion,2021_RMP_Ion,2022_Zhan_Rydberg_gate}, coherent synthesis of ultracold molecules~\cite{2008Ni_molecules,2014_Hanns_RbCs,2014_Cornish_RbCs,2016_WangDaJun_NaRb,2020_Silke_NaK,2021_LuoXinYu_NaK,2022Bo_molecules}, and search for dark matter and variations in fundamental constants~\cite{2008_Flambaum_PRL_constants,2022Masami_PRL_DarkMatter}, among others. At the heart of all these applications lies a frequency-discriminating and locking method called the Pound-Drever-Hall (PDH) technique~\cite{1983_PDH_original}, which converts the frequency deviation of a laser from cavity resonance into an electrical signal suitable for high-speed feedback~\cite{PDH_Black}. Combined with high-finesse ultra-stable optical cavities, the PDH feedback technique has now been routinely used to realize laser systems with ultra-narrow linewidth~\cite{2012_YeJun_40mHz,2008_Udem_Subhertz,2016MaLS_laser,2022_LiuKaiKai_36Hz}.
	
\begin{figure*}
		\centering
		\includegraphics[width=1.98\columnwidth]{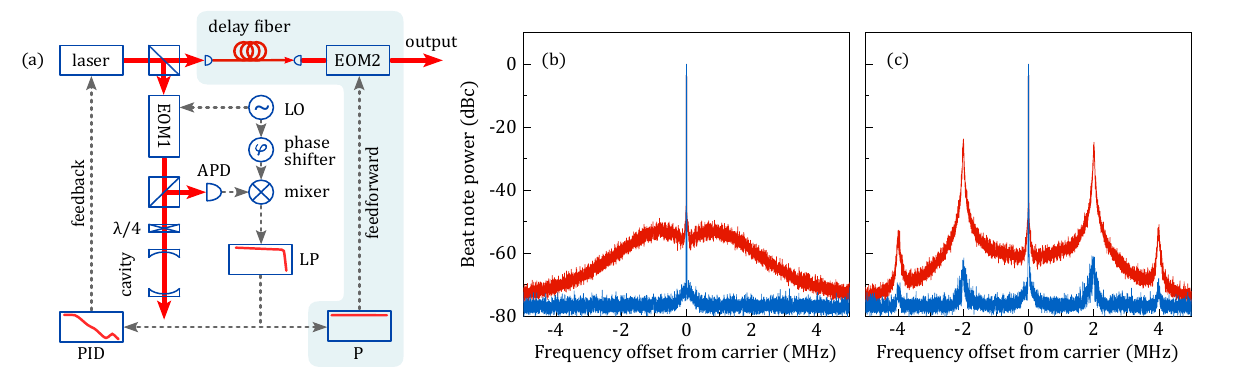}
		\caption{
			Schematics and effects of using PDH technique for both feedback and feedforward control of a light source. (a) A typical PDH feedback setup (white background) for locking the frequency of a laser to a cavity resonance is extended to suppress high-frequency phase noise using feedforward. The feedforward construction (blue shaded area) includes only a loop filter with flat gain, a delay fiber and an electro-optic modulator (EOM2). APD, avalanche photodiode. LO, local oscillator. LP, low-pass filter. (b) and (c) Power spectra of heterodyne beat between the light transmitted through the cavity and the output light after EOM2, with (blue) and without (red) applying feedforward. In (b), phase noise after feedforward is submerged under the detection noise. To reveal the full potential of PDH feedforward, the gain of the feedback loop is intentionally increased to make the servo bumps more visible in (c). A 42 dB suppression can be seen at 2 MHz. The data are recorded with a resolution bandwidth of 5 kHz. The PDH modulation frequency is 43 MHz.
		}
		\label{fig1}
	\end{figure*}

	As the name implies, however, any feedback mechanism inherently introduces a time delay $\tau$, which restricts the feedback bandwidth to $f_B \sim 1/\tau$. As a rule of thumb, noise at Fourier frequency $f_n$ can at best be suppressed by a factor of $\sim(f_B/f_n)^2$ in terms of noise power for an optimized feedback loop. Beyond $f_B$ however, the feedback loop no longer works to suppress the noise, but rather tends to amplify it, resulting in the formation of servo bumps (cf. Fig.~\ref{fig1}(b,c)). These servo-induced noise bumps occur typically around a few hundred kHz to a few MHz for most PDH applications.
	
	For optical-clock applications featuring long probing times of seconds, the MHz servo bumps are generally less of a problem. For photonic microwave generation, however, noise at short timescales (milliseconds to microseconds) is much more critical~\cite{2011Fortier_Optical_microwave,2017XieMicrowave,2020Quinlan_Science_Microwave,2020Kippenberg_photonicMicrowave}, since it is mapped directly onto the spectrum of the generated microwave. The ongoing race to build quantum computers using Rydberg atoms and trapped ions further highlights the deleterious effects of the servo-induced noise on the realization of high-fidelity quantum gates~\cite{2018_Browaeys_Coherence,2018_Levine_HighFidelity,2019_JDPritchard_CsEntanglement,2022_Senko_Limits,2023_Saffman_GateFidelity,2023_Whitlock_StandingWave,2014_Biercuk_Ion_NoiseFiltering}, particularly when the occurrence of servo bumps coincides with the Rabi frequency of the gates~\cite{2018_Sylvain_Thesis}. Untamed servo bumps can also diminish the synthesis efficiency of ultracold molecules~\cite{2021_LuoXinYu_NaK}.
	
	To address these challenges, the most widely adopted solution is to use optical cavities for spectral filtering~\cite{2018_Levine_HighFidelity,2005_JanHald_DoubleFilterCavity,2015_Akerman_Ion_FilterCavity,2021_LuoXinYu_NaK}. Nonetheless, the power transmitted through a narrow-linewidth cavity is typically limited to a few mW or less to prevent damage to the cavity coatings caused by accumulation of intense optical power within the cavity. For most applications therefore, one either has to compromise on the filtering linewidth or to use additional low-noise laser amplifiers.
	
	An alternative approach is to perform feedforward~\cite{2009_Hossein_MZI,2012_Hossein_MZI,2016_Tetsuya_TrackingInterferometer,2016_Barry_FrequencyComb,2017_Lintz_Note,2019_HeZuyuan_AOM,2022_Demarco_CavityRef,2015Scharnhorst_Feedforward,2016Xu_FeedforwardwithComb}. In contrast to feedback, the feedforward scheme uses an additional fiber segment (Fig.~\ref{fig1}(a)) to hold up the light before applying the phase-noise-compensating signal, so as to account for the delay in noise detection and feedforward process. A properly compensated feedforward can therefore be considered as a real-time feedback, and thus has a much higher bandwidth. Nevertheless, a feedforward is typically more susceptible to unforeseen perturbation than a feedback. Consequently, it is crucial to devise reliable methods for generating and applying a feedforward signal. Previous feedforward experiments have employed delayed self-homodyne measurements using an unbalanced Mach-Zehnder interferometer (MZI)~\cite{2009_Hossein_MZI,2012_Hossein_MZI,2016_Tetsuya_TrackingInterferometer,2016_Barry_FrequencyComb,2017_Lintz_Note,2017_HuangWei_CoherentEnvelope,2019_ZhanMingSheng_EnvelopRatio,2021_Polzik_HighFrequency,2017_ZhaoJianye_OpticalDelayLine}, or heterodyne measurements between a light and its transmission through a filtering cavity~\cite{2022_Demarco_CavityRef} to detect high-frequency phase noise. These methods necessitate an additional feedback loop to counteract any drifts in the optical paths between the two arms of the homodyne/heterodyne measurements. Given that the PDH signal contains all spectral information of the laser noise~\cite{2019_Udem_SimpleMeasurement,2018_Thomas_ResidualSuppression},one might contemplate whether the PDH technique, traditionally employed solely for feedback, can also be utilized for feedforward.
	
	In this work, we demonstrate the feasibility and methodology of using PDH signal for feedforward purposes. Through an in-depth analysis of the residual PDH error signal for a laser locked on a cavity, we discover that it can be directly applied to perform feedforward using a loop filter with a frequency-independent gain. By implementing our proposed approach to a common laser setup, we demonstrate a noise suppression exceeding 37~dB at 2~MHz, outperforming usual PDH feedback by four orders of magnitude and surpassing previous feedforward demonstrations by at least an order of magnitude \cite{2009_Hossein_MZI,2012_Hossein_MZI,2016_Tetsuya_TrackingInterferometer,2016_Barry_FrequencyComb,2017_Lintz_Note,2019_HeZuyuan_AOM,2022_Demarco_CavityRef,2015Scharnhorst_Feedforward,2016Xu_FeedforwardwithComb} at the specified frequency. We further show that the PDH feedforward scheme exhibits noise suppression capability on par with cavity filtering at low frequencies. The limits on its performance and the possible improvements will also be discussed.
	
	{\it Experimental setup and effects. ---}
	The schematic diagram of Fig.~\ref{fig1}(a) demonstrates how to apply both feedback and feedforward to suppress laser phase noise using PDH technique. The primary part of the construction is the familiar PDH feedback~\cite{1983_PDH_original,PDH_Black}, which consists mainly of: (1) a local oscillator (LO) and an electro-optic modulator (EOM1) for phase modulation and generation of frequency sidebands; (2) an optical cavity that imparts distinct reflections to the far-off-resonant sidebands and the near-resonant carrier; (3) an avalanche photodiode (APD) for detecting the beat note between the reflected sidebands and carrier; (4) a mixer and a low-pass filter (LP) for converting the frequency offset between the laser and a cavity resonance into an dispersive `error signal' suitable for frequency locking. For feedback, this error signal is applied to the laser after traversing a loop filter whose output is generally a summation of the proportional, integral and derivative (PID) terms of the input signal. The key message of this manuscript is that it is possible to efficiently suppress the remaining phase noise of a laser feedback-locked to a cavity by recycling the residual PDH error signal under locking condition and feedforwarding it to the laser again. The method requires only a few supplemental components: a delay fiber, an EOM for phase noise compensation, and a loop filter with a constant proportional (P) gain (as enclosed within the blue shaded region in Fig.~\ref{fig1}(a)).
	
	The phase noise suppression performance of this feedforward method is demonstrated in Figs.~\ref{fig1}(b) and (c), utilizing a 1013-nm external-cavity diode laser (ECDL), an ultra-low expansion (ULE) cavity with a FWHM linewidth of $\gamma_c/2\pi \sim14.5$~kHz, a 20-meter long delay fiber and two home-built loop filters (PID and P). The figures compare the power spectra of the beat signal between the cavity-filtered light and the output light passing through EOM2, with (blue) and without (red) feedforward (see Section~1 in supplemental material for detailed experimental setup and Section~4 in supplemental material for feedforward optimization procedures). For the plotted frequency range, the PDH-locked cavity-filtered light can be considered as an ideal frequency reference~\cite{comment_Freq_Ref}. Therefore, the heterodyne spectra reflect essentially the actual noise spectra of the output light for noise frequencies $\gg \gamma_c$. In Fig.~\ref{fig1}(b) where a moderate gain is chosen for the feedback loop, the servo bumps peak around 700~kHz with a magnitude of about -53~dBc. Upon applying the feedforward signal to EOM2, the phase noise is significantly suppressed, to a level below the detection noise of the heterodyne setup for frequencies above 500~kHz. To fully explore the potential of PDH feedforward, we intentionally increase the gain of the feedback loop, boosting the servo bumps to 2~MHz and -24~dBc (Fig.~\ref{fig1}(c)). Under this condition, a maximum 42~dB suppression is observed at the servo bumps with feedforward. This suppression ratio is an order of magnitude higher than previously reported feedforward schemes~\cite{2009_Hossein_MZI,2012_Hossein_MZI,2016_Tetsuya_TrackingInterferometer,2016_Barry_FrequencyComb,2017_Lintz_Note,2019_HeZuyuan_AOM,2022_Demarco_CavityRef,2015Scharnhorst_Feedforward,2016Xu_FeedforwardwithComb}.
	
	\begin{figure}[!h]
		\centering
		\includegraphics[width=0.98\columnwidth]{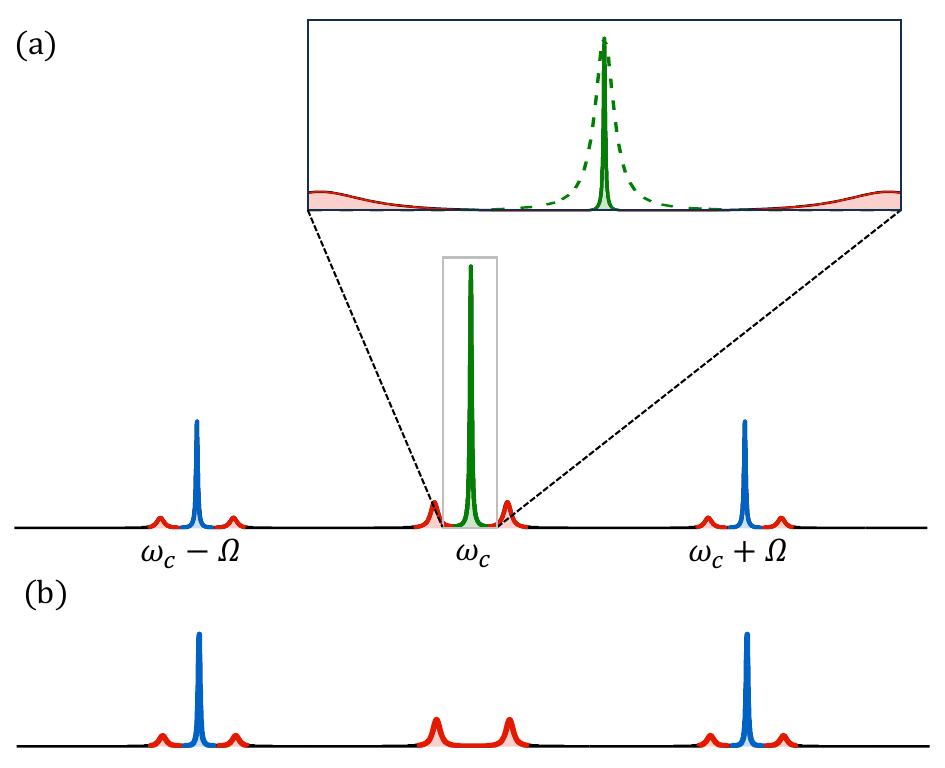}
		\caption{
			Spectra of feedback-locked laser before (a) and after (b) cavity reflection. (a) The green, blue, and red lines represent the carrier, the first order sidebands, and the residual phase noise (servo bumps), respectively. The dashed green line in the zoom-in (inset) represents the transmission spectrum of the cavity. Typically, the locked-carrier linewidth is far narrower than the cavity linewidth, while the residual servo bumps are far away from the cavity transmission window. (b) Under locking conditions, the carrier transmits through the cavity and is thus missing from the reflection spectrum.
		}
		\label{fig2}
	\end{figure}
	
	{\it Working Principles. ---}
	This section aims to establish the theoretical foundation for the implementation of PDH feedforward. First, we note that the prerequisite for this method to work is that the light source is already well-feedback-locked to the cavity. Under this condition, the typical linewidth of the locked light source is well below $\sim0.01\gamma_c$, and the remaining untamed phase noise or servo bumps are typically of characteristic frequencies $\omega_n \gg \gamma_c$ (as is illustrated by the inset of Fig.~\ref{fig2}(a)). We can therefore, to a very good approximation, express the electric field of the locked light source as $E_0 \exp \left\{i\left[ \omega_c t + \phi(t)\right]\right\}$. Here, $E_0$ is the field amplitude, $\omega_c$ denotes the resonant frequency of the cavity, and $\phi(t)$ represents the instantaneous residual phase noise at time $t$. Through EOM1, the PDH technique imprints a phase modulation of amplitude $\beta$ at angular frequency $\Omega\gg \omega_n$ to the light, resulting in a field of
	\begin{equation} \label{eq_incident}
		E_{\rm in}(t) = E_0 \exp \left\{i\left[ \omega_c t + \phi (t) + \beta \sin(\Omega t) \right]\right\},
	\end{equation}
	which can be expanded as
	\begin{equation} \label{eq_bessel}
		E_{\rm in}(t)= E_0 \sum_k J_k(\beta) \exp \left\{i\left[(\omega_c + k\Omega) t+ \phi (t)\right] \right\},
	\end{equation}
	where $J_k(x)$ represents the $k^{\rm th}$ order Bessel function of the first kind ($k=0, \pm 1, \pm 2, ...$). Equation~(\ref{eq_bessel}) signifies that the incident field before the cavity can be decomposed into a series of field components of frequencies $\omega_c + k\Omega$, each accompanied by the servo bumps as illustrated by Fig.~\ref{fig2}(a). Since $\Omega\gg \omega_n \gg \gamma_c$, we assume that all of these field components, except for the carrier at $\omega_c$, are completely reflected by the cavity. Without loss of generality, we further assume that the on-resonant carrier is fully transmitted by the cavity~\cite{realcavity}. With these assumptions, the field reflected from the cavity can be expressed by
	\begin{equation} \label{eq_reflected}
		E_{\rm ref}(t)= -E_{\rm in}(t)  + E_0 J_0(\beta) \exp(i\omega_c t),
	\end{equation}
	where the negative sign in the first term indicates a phase shift of $\pi$ acquired by the reflected fields.
	Under this simplified model, the PDH error signal (after passing through the APD, mixer and low-pass filter) is given by (see Section 2 in supplemental material for calculation details)
	\begin{equation} \label{eq_error}
		V_\mathrm{error}(t) \propto \sin\left[\phi(t)\right].
	\end{equation}
	Consequently, when a phase shift proportional to $V_\mathrm{error}(t)$ is applied to the light source through an EOM, the output field becomes $E_{\rm out} = E_0 \exp\left\{i \left[ \omega_c t + \phi(t)+ G \sin\phi(t^{\prime})\right]\right\}$. Here, a different time $t^\prime$ is used to highlight the potential delay in the feedforward signal. It is evident that, when $|\phi(t)|\ll 1$, the phase noise $\phi(t)$ can be effectively compensated by setting a constant $G=-1$ for all frequencies, and matching $t$ and $t^\prime$ (using the delay fiber in Fig.~\ref{fig1}(a)). Experimentally, the latter condition can be realized easily by first choosing a delay fiber longer than the estimated effective length~\cite{refractiveindex} of the optical and electrical paths combined (i.e. from the laser head to the APD, and then from the APD to EOM2), and then compensating the mismatch by inserting coaxial cables in the feedforward path.
	
	\begin{figure}
		\centering
		\includegraphics[width=0.98\columnwidth]{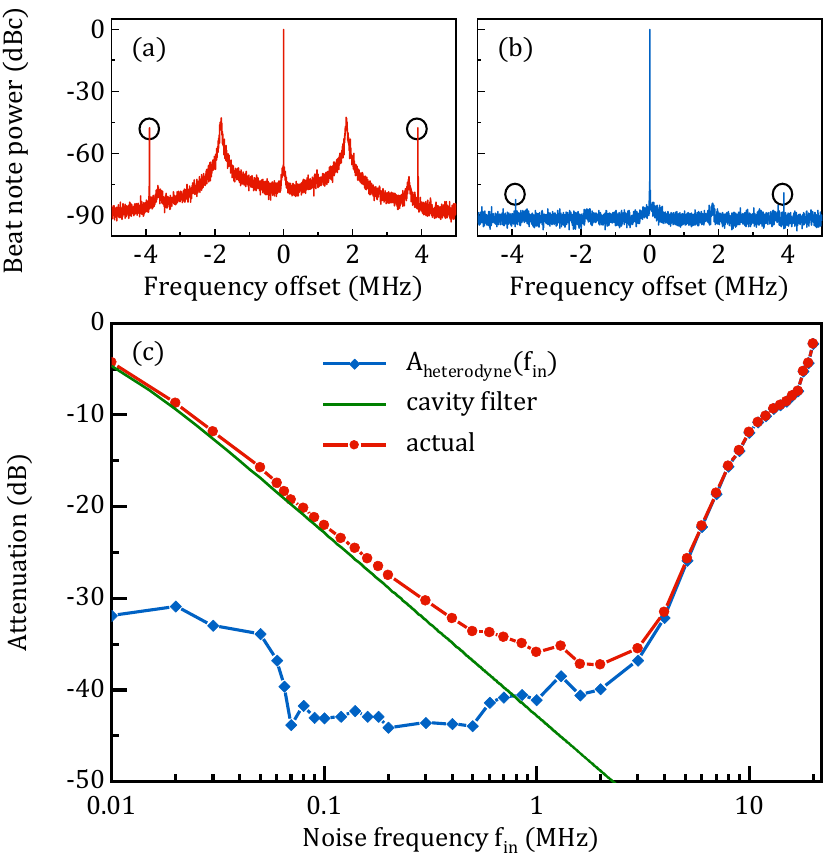}
		\caption{
			Performance and bandwidth of PDH feedforward. (a) and (b), heterodyne spectra similar to Figs.~\ref{fig1}(b,c) with artificially injected sinusoidal phase noise (black circles) before and after applying feedforward, respectively. The data are recorded with a resolution bandwidth of 200~Hz. (c) Noise attenuation performance as a function of injected noise frequency $f_{\rm in}$. The blue diamonds represent the attenuation parameter $A_{\rm heterodyne}(f_{\rm in})$ (see text) obtained from data similar to (a) and (b). The green solid line shows the noise filtering effect for the cavity we use, which is equivalent to ideal PDH feedforward (see text). The red dots indicate the conservative lower bound of the actual noise attenuation. For all measurements, the beat note powers of the injected sidebands are kept around $-47.8\pm 1.5$~dBc before applying feedforward.
		}
		\label{fig3}
	\end{figure}
	
	{\it Attenuation bandwidth. ---}
	We now investigate the noise-suppression performance of our feedforward scheme across a wider frequency range. To this end, we artificially inject a weak sinusoidal phase modulation into the light field using an additional EOM (see Section~1 in supplemental material). For all modulation frequencies $f_{\rm in}=\omega_n/2\pi$ investigated here, the weak injection causes no discernible changes to the noise spectrum of the feedback-locked laser, except for the emergence of a pair of injected sidebands, as indicated by the circles in Figs.~\ref{fig3}(a) and (b). We quantify the feedforward performance by defining an attenuation parameter $A_{\rm heterodyne}(f_{\rm in}) \equiv P_{\rm w}(f_{\rm in})/P_{\rm w/o}(f_{\rm in})$, where $P_{\rm w(w/o)}$ represents the beat note power of the injected sidebands measured from the heterodyne spectrum between the cavity-filtered field and the output laser field with (without) feedforward. The results (blue diamonds in Fig.~\ref{fig3}(c)) show a remarkable suppression of over 30~dB from 10~kHz to 4~MHz, reaching a maximum of 43~dB, before exhibiting a decline at higher frequencies. This decline is primarily attributed to the dispersion in group velocity of the low-pass filter used in our setup. Surprisingly, the measured phase-noise attenuation at low frequencies appears to surpass the performance of cavity filtering (green solid line).
	
	{\it Performance limits. ---}
	To comprehend the measured data in Fig.~\ref{fig3}, we need to expand the previous model to account for the cavity's response to lower frequency noise. Considering a weak single-frequency phase noise $\phi^n_{\rm in}(t)= \beta_n \sin(\omega_n t)$ ($|\beta_n|\ll1$) and referring to the schematic in Fig.~\ref{fig2}, we now posit that the cavity completely reflects all field components, except for the three at $\omega_c-\omega_n$, $\omega_c$, and $\omega_c+\omega_n$. This approximation results in a PDH error signal
	\begin{equation}\label{eq_PDH_error_low_freq}
		V_{\rm error}^{(2)}(t) \propto\mathrm{Re}\mathcal{R}(\omega_n)\sin(\omega_n t) + \mathrm{Im}\mathcal{R}(\omega_n)\cos(\omega_n t),
	\end{equation}
	where $\mathrm{Re}\mathcal{R}(\omega_n)$ and $\mathrm{Im}\mathcal{R}(\omega_n)$ represent the real and imaginary parts of the cavity reflectivity $\mathcal{R}(\omega_n)$ (Fig.~\ref{fig4}(a))~\cite{realcavity} (see Section~3 in supplemental material for calculation details). Compared to Eq.~(\ref{eq_error}), this error signal contains an additional quadrature term $\mathrm{Im}\mathcal{R}(\omega_n)\cos(\omega_n t)$ which diminishes for large $\omega_n$. Feedforwarding this error signal to the light source by setting $t=t^\prime$ and choosing a gain that eliminates all high-frequency phase noise results in a suppressed phase noise of
	\begin{equation}\label{FF_phase_noise}
		\phi^n_{\rm out}(t)=\beta_n\left\{\left[1+ \mathrm{Re}\mathcal{R}(\omega_n)\right] \sin(\omega_n t) + \mathrm{Im}\mathcal{R}(\omega_n) \cos(\omega_n t)\right\}.
	\end{equation}
	This result signifies that the effect of PDH feedforward is equivalent to cavity filtering, since $1+ \mathcal{R}(\omega_n)=\mathcal{T}(\omega_n)$ where $\mathcal{T}(\omega_n)$ is the complex field transmissivity of the cavity.
	
	\begin{figure}
		\centering
		\includegraphics[width=1\columnwidth]{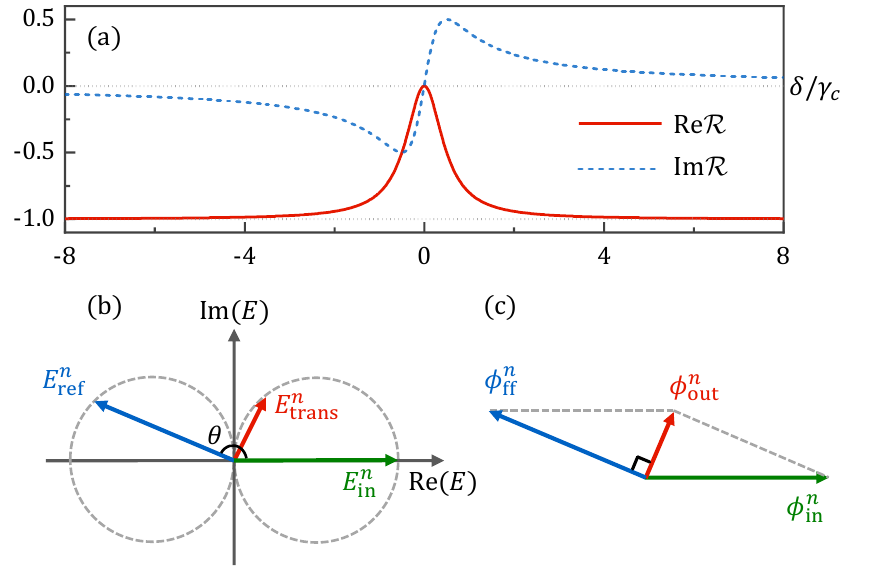}
		\caption{
			Working principle of PDH feedforward. (a) Complex field reflectivity $\mathcal{R}(\delta)$ of a symmetric cavity with no loss, where $\delta$ is the frequency offset from the cavity resonance. (b) Relationships among the cavity's incident $E^n_{\rm in}$, reflected $E^n_{\rm ref}$, and transmitted $E^n_{\rm trans}$ fields (see text). (c) The input phase noise $\phi^n_{\rm in}$, the feedforward phase $\phi^n_{\rm ff}$, and the feedforward-suppressed phase noise $\phi^n_{\rm out}$ inherit the relative amplitudes and angles among the respective fields in (b) through PDH feedforward. $\phi^n_{\rm out} = \phi^n_{\rm in}+\phi^n_{\rm ff} = \phi^n_{\rm trans}$ (Eq.~(\ref{FF_phase_noise})) vanishes for $\delta\gg\gamma_c$.
		}
		\label{fig4}
	\end{figure}
	
	For better understanding of the above conclusion, it is illuminating to visually represent the relationships among the fields of cavity's incidence, reflection and transmission on a complex plane (Fig.~\ref{fig4}(b)). Consider an incident (noise) field component $E^n_{\rm in}$ at frequency $\omega_c+\delta$, which is positioned along the real axis as a reference. For a critically coupled cavity, the reflected field $E^n_{\rm ref}$ traces out a circle with diameter |$E^n_{\rm in}$| intersecting the origin~\cite{PDH_Black,realcavity}. The phase shift between $E^n_{\rm in}$ and $E^n_{\rm ref}$ is $\theta=\pi/2+\tan^{-1}(2\delta/\gamma_c)$. Starting from the origin when $\delta=0$, $E^n_{\rm ref}$ traces the circle counter-clockwise as $\delta$ increases, and points to the opposite of $E^n_{\rm in}$ when $\delta\gg\gamma_c$. Similarly, the transmitted noise field, $E^n_{\rm trans}$, also traces out a circle, but lagging $E^n_{\rm ref}$ by a phase shift of $\pi/2$. Note that $E^n_{\rm trans}=E^n_{\rm in}+E^n_{\rm ref}$ for all $\delta$. There exists a direct mapping from the aforementioned cavity fields to the feedforward phases (Fig.~\ref{fig4}(c)). The faithful mapping of the amplitude and phase of $E^n_{\rm ref}$ onto the PDH signal (Eq.~(\ref{eq_PDH_error_low_freq})), together with the feedforward requirements of setting $t=t^\prime$ and $G=-1$, effectively transfers the relative amplitudes and angles inherent to $E^n_{\rm ref}$, $E^n_{\rm in}$ and $E^n_{\rm trans}$, to the feedforward phase $\phi^n_{\rm ff}$, the input phase $\phi^n_{\rm in}$ and the output phase $\phi^n_{\rm out}$. The correspondence between $\phi^n_{\rm out}$ and $E^n_{\rm trans}$ means that laser-phase-noise suppression using PDH feedforward is equivalent to cavity filtering.
	
	When amplitude noise is negligible, the noise spectrum of heterodyne detection reflects the difference in the phase noise between the two interfering light fields, rather than the summation of them. Consequently, if the two beating beams have identical phase noise, the power spectrum of heterodyne detection show no noise (see Section~3 in supplemental material for a more detailed discussion). Revisiting Fig.~\ref{fig3}(c), since perfect PDH feedforward has the same effect as cavity filtering, $A_{\rm heterodyne}(f_{\rm in})$ should approaches 0 in the ideal situation. The observed non-zero values (blue diamonds in Fig.~\ref{fig3}(c)) are attributed to experimental imperfections. As is shown in Section~3 of supplemental material, the measured  $A_{\rm heterodyne}(f_{\rm in})$ is related to the difference between the actual (with imperfect feedforward) output phase noise $\phi^{n}_{\rm out, im}$ and the cavity transmission phase noise $\phi^{n}_{\rm trans}$ by $A_{\rm heterodyne}(f_{\rm in}) =|\phi^{n}_{\rm out, im}-\phi^{n}_{\rm trans}|^2/|\phi^{n}_{\rm in}|^2=|\phi^n_{\rm beat}|^2/|\phi^{n}_{\rm in}|^2$. With this understanding, we can compute a conservative lower bound for the actual noise attenuation (red dots in Fig.~\ref{fig3}(c)) by assuming that the remaining phase noise amplitude in the feedforwarded output field $|\phi^{n}_{\rm out, im}|=|\phi^{n}_{\rm trans}|+|\phi^n_{\rm beat}|$ (see Eq.~(S20) in supplemental material). The results show that the feedforward performance closely tracks the cavity filtering limit up to about 1~MHz. At higher frequencies, imperfections in the experimental settings cause the feedforward scheme to under-perform cavity filtering. In this case, the actual attenuation aligns with $A_{\rm heterodyne}(f_{\rm in})$ as expected.
	
	Note that, to achieve attenuation better than 40~dB, one needs a gain deviation from $G=-1$ of less than 0.01 (see Section~5 in supplemental material). Such a gain flatness is achievable from DC to several tens of MHz by some high-speed amplifiers, including commercial amplifiers with fixed or variable gain. Also, the performance at higher frequency can be improved by choosing low-pass filters with higher cut-off frequency. 
	
	Attention should be paid to several aspects when applying PDH feedforward. Firstly, the amplitude of PDH error signal varies with the cavity transmission power. When transmission drifts, the overall feedforward gain $G$ deviates from $-1$ and the feedforward performance becomes worse. Therefore, we recommend stabilizing the power of the cavity transmission to improve its long-term stability for the more demanding applications (see Section~5 in supplemental material). Secondly, the delay fiber can introduce low-frequency ($\lesssim$ kHz) phase noise into the laser when it experiences mechanical vibration and temperature drift. Therefore, it would be helpful to use a faster feedforward circuitry to reduce delay, and also to avoid using a fiber much longer than necessary. For applications which are sensitive to low frequency noise, there exists established technique for canceling phase noise introduced by optical fibers~\cite{1994_LongShengma_DeliverFrequency}. Thirdly, the feedforward EOM (EOM2) may inadvertently introduce amplitude modulation into the light field when performing phase modulation, but such side effect should be small in general if attention is paid to avoid substandard EOM and bad optical alignment. In our setup, the induced power fluctuation is measured to be below $0.04\%$ (root-mean-square) of the total power when PDH feedforward is applied.
	
	{\it Summary and outlook. ---}
	We show that the residual PDH error signal of a laser locked to a cavity is a direct map of its instantaneous phase noise, and therefore can be straightforwardly fed forward to that laser again to eliminate high-frequency phase noise. By combining feedback and feedforward, the PDH method now offers unparalleled suppression of laser phase noise from DC up to tens of MHz. Compared to previous feedforward schemes~\cite{2009_Hossein_MZI,2012_Hossein_MZI,2016_Tetsuya_TrackingInterferometer,2016_Barry_FrequencyComb,2017_Lintz_Note,2019_HeZuyuan_AOM,2022_Demarco_CavityRef,2015Scharnhorst_Feedforward,2016Xu_FeedforwardwithComb}, the PDH feedforward approach obviates the need for additional homodyne or heterodyne measurements for probing phase noise, and is therefore more robust against optical path variations and intensity noise. Furthermore, since PDH feedforward has the same noise suppression capability equivalent to cavity filtering but does not suffer from limited power of cavity transmission, one may choose a cavity with a minimal linewidth for optimal noise suppression at low frequencies. The new method holds great promise for applications demanding highly stable lasers with significant power output and attenuated phase noise up to tens of MHz.

	\begin{acknowledgments}
		{\it Acknowledgements. ---} This work is supported by the National Natural Science Foundation of China (NSFC) (Grant No. 12234012 and No. 92265205), the National Key R\&D Program of China (Grant No. 2018YFA0306503), and the Innovation Program for Quantum Science and Technology (2021ZD0302104).
	\end{acknowledgments}

	\bibliographystyle{apsrev4-2}
	
	%

\pagebreak
\clearpage
\onecolumngrid

\setcounter{equation}{0}
\setcounter{figure}{0}
\setcounter{table}{0}
\setcounter{page}{1}
\renewcommand{\theequation}{S\arabic{equation}}
\renewcommand{\thefigure}{S\arabic{figure}}
\renewcommand{\bibnumfmt}[1]{[S#1]}

\begin{center}
	\textbf{\large {Supplemental Material for \\
			``Pound-Drever-Hall Feedforward: Laser Phase Noise Suppression beyond Feedback''}}
\end{center}

This supplemental material contains details regarding the experimental setup,  calculations of PDH error signal, the procedures for optimizing feedforward, stability of PDH feedforward, and how we obtain the conservative lower bound for the feedforward phase-noise suppression.

\section{Experimental setup}
\begin{figure}[!htp]
	\centering
	\includegraphics[width=0.55\columnwidth]{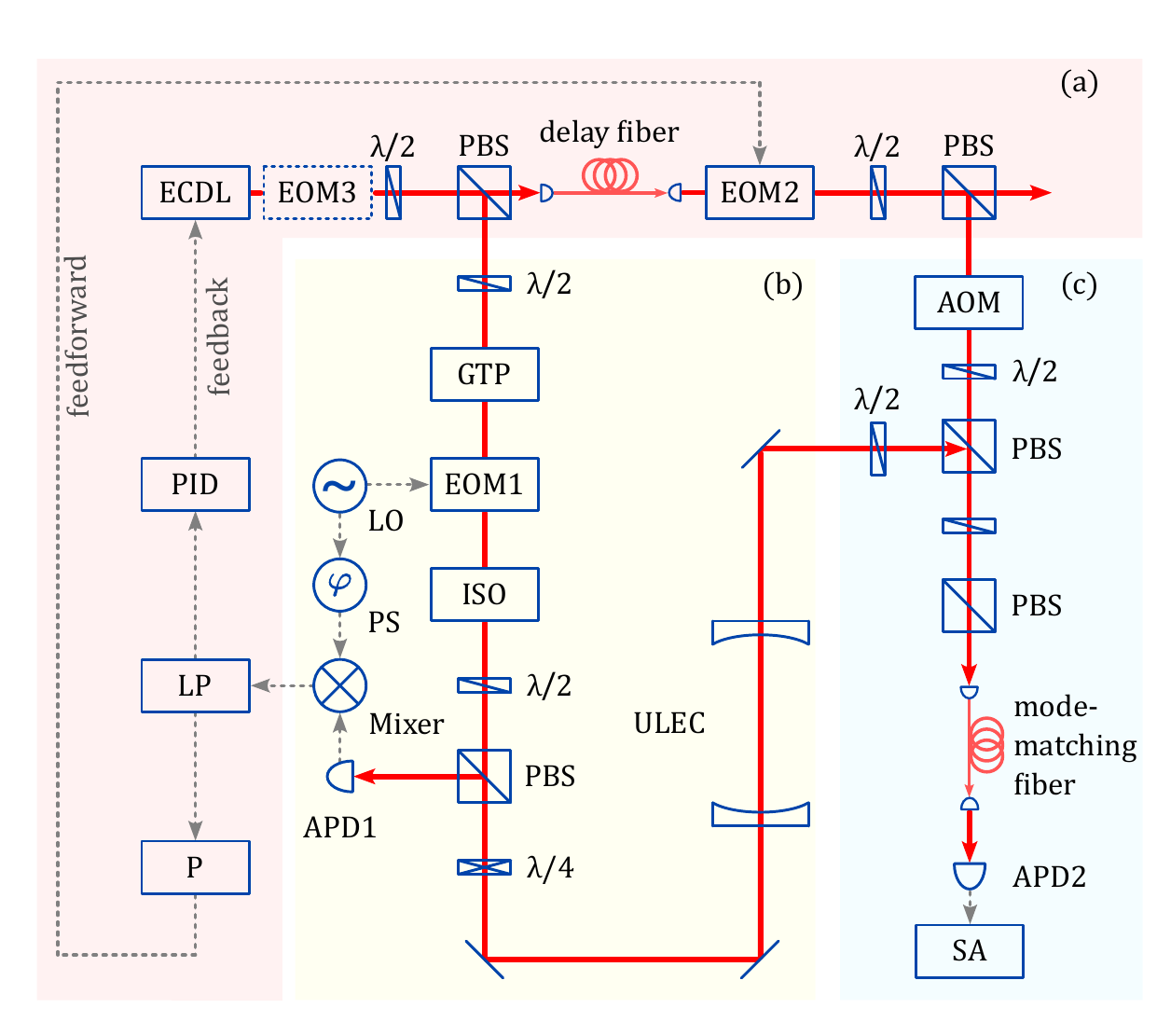}
	\caption{
		Detailed schematic of our experimental setup. Part (a) shows the feedback  and feedforward control systems; part (b) works to generate the PDH error signal; part (c) is the setup for measuring the self-heterodyne spectrum. ECDL, external-cavity diode laser. EOM, electro-optic modulator for phase modulation). PBS, polarizing beam splitter. LP, low-pass filter. PID, loop filter whose output is the sum of proportion, integral, and derivative of the input signal. P, loop filter whose output is proportional to the input signal. GTP, Glan-Taylor polarizer. ISO, isolator. LO, local oscillator. PS, phase shifter. APD, avalanche photodiode. ULEC, cavity made of ultra-low expansion glass. AOM, acousto-optic modulator. SA, RF spectrum analyzer.
	}\label{s1}
\end{figure}

Figure~\ref{s1} reveals more details of our experimental setup. The light source in the setup is a 1013-nm external cavity diode laser (ECDL, Toptica DL pro). A high-finesse cavity with a FWHM linewidth of 14.5~kHz  is used to implement the Pound-Drever-Hall (PDH) locking using an incident power of 150~\si{\micro\watt} before the cavity. A home-made electro-optic modulator (EOM1) is used to produce a 43-MHz phase modulation with a modulation index $\beta \sim 1.08$. The light reflected from the cavity is detected by an avalanche photodiode (APD1, Thorlabs APD430C). After demodulation, we use a low-pass filter (LP, Minicircuits SLP21.4+) to eliminate high-frequency ($\gtrsim 20$\,MHz) noise in the error signal. One half of PDH error signal is fed back to control the current of the ECDL, thereby locking laser frequency to the cavity resonance; another half is fed forward to a fiber EOM (EOM2, JENOPTIK PM1064) after passing through a home-made loop filter with a flat gain (P). 

The self-heterodyne spectra in Fig.~1(b,c) and Fig.~3(a,b) in the main text are measured using the setup shown in Fig.~\ref{s1}(c). We use an acousto-optic modulator (AOM) to shift the frequency of feedforwarded light after EOM2, and overlap it with the cavity transmitted light using a polarizing beam splitter (PBS). To allow beating between the two orthogonal polarized light after the PBS, we use a half-wave plate to rotate the polarization of the two beams and then use another PBS to pick out field components with the common polarization. After that, both beams are coupled into a single-mode fiber to ensure perfect mode matching. The beat signal is detected by another APD (APD2, Thorlabs APD430C), whose output goes to a spectrum analyzer (SA, Rohde $\&$ Schwarz FSV3000). The powers of the cavity-filtered light and the feedforwarded light are about 3~\si{\micro\watt} and 25~\si{\micro\watt} before APD2, respectively. The beat note power of two carriers is about 6~dBm.

To obtain the data of Fig.~3 in the main article, we place an EOM (EOM3, KEYANG PHOTONICS KY-PM-10-10G) after the laser source (Fig.~\ref{s1}(a)) to inject a weak phase modulation ($<-45$~dBc) to imitate phase noise of variable frequency.

\section{Residual PDH error signal of the simplified model}
In this section, we would calculate the residual PDH error signal of the simplified model discussed in the main text. Our feedforward scheme works under the condition that the laser is already feedback-locked to a cavity resonance at $\omega_c$. Here we assume weak residual phase noise $\phi(t)$ ($|\phi(t)|\ll1$) with Fourier frequency $\omega_n \gg \gamma_c$ ($\gamma_c/2\pi$ being the FWHM linewidth of the cavity). The incident field before the optical cavity can be expressed as~\cite{1983_PDH_original,PDH_Black}
\begin{equation} \label{eq_bessel_SM}
	E_{\rm in}(t)= E_0 e^{i\left[\omega_c t+ \beta \sin(\Omega t)+\phi (t)\right]}= E_0 \sum_k J_k(\beta) e^{i\left[(\omega_c + k\Omega) t+ \phi (t)\right]}.
\end{equation}
Here, $E_0$, $\Omega$, and $\beta$ represent the amplitude of electric field, the PDH modulation frequency $\gg \omega_n$, and the PDH modulation amplitude, respectively. $J_k(x)$ represents the $k^{\rm th}$ order Bessel function of the first kind ($k=0, \pm 1, \pm 2, \cdots$). Since $\Omega\gg\omega_n\gg\gamma_c$, we assume that all frequency components are reflected completely, except for the carrier which we assume to be fully transmitted. Under these approximations, the reflected field after the cavity is given by
\begin{equation} \label{eq_reflected_SM}
	E_{\rm ref}(t)= -E_0 \sum_k J_k(\beta) e^{i\left[(\omega_c + k\Omega) t+ \phi (t)\right]} + E_0 J_0(\beta) e^{i\omega_c t}.
\end{equation}
APD1 detects power of the reflected light $\propto |E_{\rm ref}|^2$ in such a way that all the high-frequency terms (non-zero order of $\omega_c$) would be averaged to constants. This gives a signal before mixer of (after removing the inconsequential constants)
\begin{eqnarray} \label{eq_PD}
	V_{\rm AC, mixer}(t) &\propto& -E_0^2J_0(\beta) \cos\left[\beta \sin(\Omega t) + \phi (t) \right]\nonumber \\
	&=&-E_0^2J_0(\beta) \left\{\cos\left[\beta \sin(\Omega t)\right]\cos\phi (t)- \sin\left[\beta \sin(\Omega t)\right]\sin\phi (t)\right\}.
\end{eqnarray}
Since demodulation with the quadrature $\sin(\Omega t)$ and filtering out high-frequency terms obeys the following relations:
\begin{equation} \label{eq_DM}
	\begin{split}
		\sin \left[ \beta \sin(\Omega t) \right] &= \sum_k J_{k}(\beta) \sin(k\Omega t)  \stackrel{demodulation}{\longrightarrow} J_1(\beta),
		\\ \cos \left[ \beta \sin(\Omega t) \right] &= \sum_k J_{k}(\beta) \cos(k\Omega t)  \stackrel{demodulation}{\longrightarrow} 0,
	\end{split}
\end{equation}
the residual PDH error signal is given by
\begin{equation} \label{eq_PDH_error}
	V_{\rm error}(t) \propto E_0^2J_0(\beta) J_1(\beta) \sin\left[\phi(t)\right].
\end{equation}

\section{Residual PDH error signal for low-frequency phase noise}

In this section, we would extend the simplified model to account for partial transmission of the cavity to low-frequency noise. Without loss of generality, we now consider a weak single-frequency phase noise.

\subsection{For $\phi_{\rm in}^n(t)= \beta_n \sin(\omega_n t),\ |\beta_n| \ll 1$}

First, we consider the phase noise of the form $\phi_{\rm in}^n(t)= \beta_n \sin(\omega_n t)$, where $\beta_n$ is a real number with magnitude $|\beta_n| \ll 1$. In this case, the incident field before the cavity is given by
\begin{eqnarray} \label{eq_incident1}
	E_{\rm in}(t) &=& E_0 e^{i \left[ \omega_c t + \beta \sin(\Omega t) + \beta_n \sin(\omega_n t) \right]}\nonumber\\
	&=& E_0 \sum_k \sum_m J_k(\beta)J_m(\beta_n) e^{i (\omega_c + k\Omega + m \omega_n) t}.
\end{eqnarray}
As $\Omega \gg \gamma_c$ remains valid, we assume that the cavity completely reflects all frequency components, except for the three at $\omega_c - \omega_n$, $\omega_c$, and $\omega_c + \omega_n$. Under this approximation, the complex reflected field can be expressed as
\begin{equation} \label{eq_reflected1}
	\begin{split}
		E_{\rm ref}(t) \approx & -E_0 e^{i \left[ \omega_c t + \beta \sin(\Omega t) + \beta_n \sin(\omega_n t) \right]}
		\\& + J_0(\beta) J_0(\beta_n) E_0 e^{i \omega_c t}
		\\& + J_0(\beta) J_1(\beta_n) [1 + {\rm Re}\mathcal{R}(\omega_n) + i{\rm Im}\mathcal{R}(\omega_n)] E_0 e^{i (\omega_c + \omega_n) t}
		\\& + J_0(\beta) J_{-1}(\beta_n) [1 + {\rm Re}\mathcal{R}(\omega_n) - i{\rm Im}\mathcal{R}(\omega_n)] E_0 e^{i (\omega_c - \omega_n) t},
	\end{split}
\end{equation}
where $\mathcal{R}(\delta)$ is the complex reflectivity of the cavity at detuning $\delta$ from the resonance, and we use the relationship $\mathcal{R}(\delta) = \mathcal{R}^*(-\delta)$ to simplify the right-hand side of Eq.~(\ref{eq_reflected1}). The signal detected by APD1 is proportional to $\left| E_{\rm ref} \right| ^2$, giving a signal before the mixer of
\begin{equation} \label{eq_PD1}
	\begin{split}
		V_{\rm AC, mixer}^{(2)}(t) \sim & E_0^2 \left\{-J_0(\beta) J_0(\beta_n) \cos \left[ \beta_n \sin(\omega_n t) + \beta \sin(\Omega t) \right]\right.
		\\&\left. - 2J_0(\beta) J_1(\beta_n) \left\{ {\rm Im}\mathcal{R}(\omega_n) \cos(\omega_n t) + [1 + {\rm Re}\mathcal{R}(\omega_n)] \sin(\omega_n t) \right\} \sin \left[ \beta_n \sin(\omega_n t) + \beta \sin(\Omega t) \right]\right\}.
	\end{split}
\end{equation}
Since $|\beta_n| \ll 1$, we can keep only the terms up to the first order of $\beta_n$. Demodulating with $\sin(\Omega t)$ and using Eq.~(\ref{eq_DM}) again, the PDH error signal is simplified as
\begin{equation} \label{eq_PDH_Error1}
	\begin{split}
		V_{\rm error}^{(2)}(t) & \sim - E_0^2J_0(\beta)J_1(\beta) \beta_n \left[ {\rm Re}\mathcal{R}(\omega_n)\sin(\omega_n t) + {\rm Im}\mathcal{R}(\omega_n)\cos(\omega_n t) \right]
		\\& = - E_0^2J_0(\beta)J_1(\beta) \beta_n \left|\mathcal{R}(\omega_n)\right| \sin(\omega_n t + \psi_{\omega_n}),
	\end{split}
\end{equation}
where $\psi_{\omega_n} = \arctan[{\rm Im}\mathcal{R}(\omega_n)/{\rm Re}\mathcal{R}(\omega_n)]$ is the phase shift acquired through the reflection process. From Eq.~(\ref{eq_PDH_Error1}), we can conclude that the PDH error signal contains both the amplitude and phase of the reflected noise field.

Finally, applying a signal proportional to $V_{\rm error}^{(2)}$ to EOM2 adds a phase shift of
\begin{equation} \label{eq_feedforward_phase_shift}
	G \beta_n \left[ {\rm Re}\mathcal{R}(\omega_n)\sin(\omega_n t') + {\rm Im}\mathcal{R}(\omega_n) \cos(\omega_n t') \right]
\end{equation}
to the output field ($G$ is the gain of the feedforward loop, $t'$ is used to highlight the possible delay in the feedforward signal). When $G=-1$ and $t'=t$, the output field after feedforward becomes
\begin{equation} \label{eq_AfterFeedforward1}
	\begin{split}
		E_{\rm out}(t) \propto & \ e^{i \omega_c t} e^{i \beta_n \left[ \sin(\omega_n t) + {\rm Re}\mathcal{R}(\omega_n)\sin(\omega_n t) + {\rm Im}\mathcal{R}(\omega_n)\cos(\omega_n t)  \right] }	
		\\ = & \ e^{i \omega_c t} e^{i \beta_n \left[ [1+{\rm Re}\mathcal{R}(\omega_n)]\sin(\omega_n t) + {\rm Im}\mathcal{R}(\omega_n)\cos(\omega_n t)  \right] },
	\end{split}
\end{equation}
which is equivalent to Eq.~(6) in the main text.

\subsection{For $\phi_{\rm in}^n(t)= \beta_n \cos(\omega_n t),\ |\beta_n| \ll 1$}

For completeness, we now show that the above conclusions are intact if we consider a phase noise in a different quadrature, i.e. $\phi_{\rm in}^n(t)= \beta_n \cos(\omega_n t)$. In this case, the complex electric field of the incident light to the cavity is
\begin{equation} \label{eq_incident2}
	E_{\rm in}(t) = E_0 e^{i \left[ \omega_c t + \beta \sin(\Omega t) + \beta_n \cos(\omega_n t) \right]},
\end{equation}
which can be transformed to
\begin{equation} \label{eq_incident2_transform}
	\begin{split}
		E_{\rm in}(t) =& \ E_0 e^{i \left\lbrace \omega_c t + \beta \sin(\Omega t) - \beta_n \sin \left[ \omega_n (t-\frac{\pi}{2\omega_n})\right]  \right\rbrace}
		\\=& \ E_0 \sum_k \sum_m J_{k}(\beta) J_{m}(-\beta_n) e^{i\left[ (\omega_c + k\Omega)t + m\omega_n(t-\frac{\pi}{2\omega_n})\right] }.
	\end{split}
\end{equation}
Following the same assumptions where only frequencies $\omega_c - \omega_n$, $\omega_c$, and $\omega_c + \omega_n$ are partly/completely transmitted by the cavity, the reflected field after the cavity is
\begin{equation} \label{eq_reflected2}
	\begin{split}
		E_{\rm ref}(t) \approx & -E_0 e^{i \left[ \omega_c t + \beta \sin(\Omega t) + \beta_n \cos(\omega_n t) \right]}
		\\& + J_0(\beta) J_0(-\beta_n) E_0 e^{i \omega_c t}
		\\& + J_0(\beta) J_1(-\beta_n) \left[ 1 + {\rm Re}\mathcal{R}(\omega_n) + i{\rm Im}\mathcal{R}(\omega_n)\right]  E_0 e^{i (\omega_c + \omega_n) t - \frac{\pi}{2}}
		\\& + J_0(\beta) J_{-1}(-\beta_n) \left[ 1 + {\rm Re}\mathcal{R}(\omega_n) - i{\rm Im}\mathcal{R}(\omega_n)\right]  E_0 e^{i (\omega_c - \omega_n) t + \frac{\pi}{2}}.
	\end{split}
\end{equation}
After demodulation with the same $\sin(\Omega t)$ and keeping the terms up to the first order of $\beta_n$, the low-passed error signal is
\begin{equation} \label{eq_PDH_Error2}
	\begin{split}
		V_{\rm error}^{(2)}(t) & \sim - E_0^2J_0(\beta)J_1(\beta) \beta_n \left[ {\rm Re}\mathcal{R}(\omega_n)\cos(\omega_n t) - {\rm Im}\mathcal{R}(\omega_n)\sin(\omega_n t) \right]
		\\& = - E_0^2J_0(\beta)J_1(\beta) \beta_n \left|\mathcal{R}(\omega_n)\right| \cos(\omega_n t + \psi_{\omega_n}).
	\end{split}
\end{equation}
Comparing Eq.~(\ref{eq_PDH_Error2}) with Eq.~(\ref{eq_PDH_Error1}), we can see that phase noises of opposite quadrature give the same relative amplitude and phase shift after PDH detection.

\subsection{Relationship between the feedforwarded field and the cavity-transmitted field}

As is proven in the last two subsections, the result of feedforward is independent of the noise quadrature. Consequently, we take $\phi_{\rm in}^n(t)= \beta_n \sin(\omega_n t)$ as an example. The electric field of transmitted light is
\begin{eqnarray} \label{eq_Trans}
	E_{\rm trans}(t) &\approx& \ E_0 J_0(\beta) J_0(\beta_n) e^{i \omega_c t}\nonumber
	\\&& + E_0 J_0(\beta) J_1(\beta_n) [{\rm Re}\mathcal{T}(\omega_n) + i{\rm Im}\mathcal{T}(\omega_n)] E_0 e^{i (\omega_c + \omega_n) t}\nonumber
	\\&& + E_0 J_0(\beta) J_{-1}(\beta_n) [{\rm Re}\mathcal{T}(\omega_n) - i{\rm Im}\mathcal{T}(\omega_n)] E_0 e^{i(\omega_c - \omega_n) t}\nonumber
	\\&\approx& \ E_0 J_0(\beta) e^{i \omega_c t} e^{i \beta_n \left[ {\rm Re}\mathcal{T}(\omega_n) \sin(\omega_n t) + {\rm Im}\mathcal{T}(\omega_n) \cos(\omega_n t)\right] },
\end{eqnarray}
where $\mathcal{T}(\delta)$ is the complex transmissivity of the cavity at detuning $\delta$ from the resonance, and we use the relationship $\mathcal{T}(\delta) = \mathcal{T}^*(-\delta)$ to simplify the right-hand side of Eq.~(\ref{eq_Trans}). Comparing Eq.~(\ref{eq_Trans}) with Eq.~(\ref{eq_AfterFeedforward1}) and using the relationship between reflectivity and transmissivity ($1 + {\rm Re}\mathcal{R} = {\rm Re}\mathcal{T}$, ${\rm Im}\mathcal{R} = {\rm Im}\mathcal{T}$), we deduce that $E_{\rm out}(t) \propto E_{\rm trans}(t)$. This is why our PDH feedforward scheme has the same noise suppression capability equivalent to cavity filtering.

\subsection{Estimation of actual attenuation parameter by self-heterodyne measurement}

\begin{figure}[h!]
	\centering
	\includegraphics[width=0.35\columnwidth]{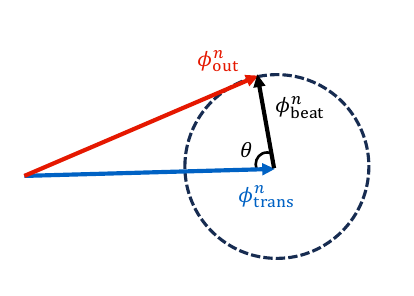}
	\caption{
		Relationship among the phase noise of the beat signal $\phi_{\rm beat}^n$, and those of the feedforwarded light $\phi_{\rm out}^n$ and cavity-transmitted light $\phi_{\rm trans}^n$ in the complex plane. $\phi_{\rm trans}^n$ can be obtained from Eq.~(\ref{eq_Trans}). The heterodyne measurement gives the amplitude of $\phi_{\rm beat}^n$. As $\phi_{\rm out}^n = \phi_{\rm trans}^n + \phi_{\rm beat}^n$, the end point of $\phi_{\rm out}^n$ can be any point on the dashed circle. Therefore, the maximum possible amplitude of $\phi_{\rm out}^n$ is $|\phi_{\rm trans}^n| + |\phi_{\rm beat}^n|$.
	}\label{fig_beat}
\end{figure}

In the last subsection, we have proven that, theoretically, PDH feedforward scheme has the same noise suppression capability equivalent to cavity filtering. However, any imperfections in applying feedforward signal would worsen its performance. Such defects include mismatch between $t$ and $t'$, deviation of the open loop gain from $G=-1$, extra noise introduced by photo-diode or other electronics, etc. The self-heterodyne beat signal in our work actually reflects the difference of the phase noise between the feedforwarded light field $E_{\rm out}$ and that of the cavity-transmitted light field $E_{\rm trans}$. Here we explain how we obtain the conservative lower bound of the actual noise attenuation presented by the red dots in Fig.~3(c) of the main text.

To emphasize the possible difference between the cavity-transmitted light and the feedforwarded light, we can write their electric field respectively as (still consider single-frequency noise of $\phi_{\rm in}^n(t)= \beta_n \sin(\omega_n t)$)
\begin{eqnarray} \label{eq_TwoField}
	E_{\rm trans}(t) &=& \ E_1 e^{i (\omega_c t + \phi_{\rm trans}^n)} = \ E_1 e^{i \left[\omega_c t + \beta_{\rm trans} \sin(\omega_n t)\right]},\\
	E_{\rm out}(t) &=& \ E_2 e^{i \left[(\omega_c+\omega_0)t + \phi_{\rm out}^n\right]} = \ E_2 e^{i \left[(\omega_c + \omega_0)t + \beta_{\rm out} \sin(\omega_n t + \psi_{\rm diff})\right]},
\end{eqnarray}
where $\omega_0$ is the frequency difference between the feedforwarded light and the transmitted light, $E_1$ and $E_2$ are the field amplitudes, $\beta_{\rm trans}$ and $\beta_{\rm out}$ are the amplitudes of phase noise, $\psi_{\rm diff}$ represents the phase difference between their phase noise. The beat of the two fields is detected by APD2 in Fig.~\ref{s1} as
\begin{eqnarray} \label{eq_Beat} 
	V_{\rm beat}(t) &\propto& \frac{1}{2}(E_1^2 + E_2^2) + E_1 E_2\cos\left[\omega_0 t -\beta_{\rm trans}\sin(\omega_n t) + \beta_{\rm out}\sin(\omega_n t + \psi_{\rm diff})\right] \nonumber \\
	&=& \frac{1}{2}(E_1^2 + E_2^2) + E_1 E_2\cos\left[\omega_0 t + \phi_{\rm beat}^n \right],
\end{eqnarray}
after removing terms of light frequency $\omega_c$ which cannot be detected by the photodetector. The spectrum of  $V_{\rm beat}$ thereby exhibits a main carrier at $\omega_0$, and two sidebands at $\omega_0 \pm \omega_n$. The phase noise in beat note is  $\phi_{\rm beat}^n = -\beta_{\rm trans}\sin(\omega_n t) + \beta_{\rm out}\sin(\omega_n t + \psi_{\rm diff}) \equiv \beta_{\rm beat} \sin(\omega_n t + \theta)$, which is just the phase noise difference between the two beating components. The relationship among $\phi_{\rm beat}^n$, $\phi_{\rm trans}^n$, $\phi_{\rm out}^n$ can be best understood by visual representation in complex plane (Fig.~\ref{fig_beat}).

\begin{figure}[htbp]
	\centering
	\includegraphics[width=0.8\columnwidth]{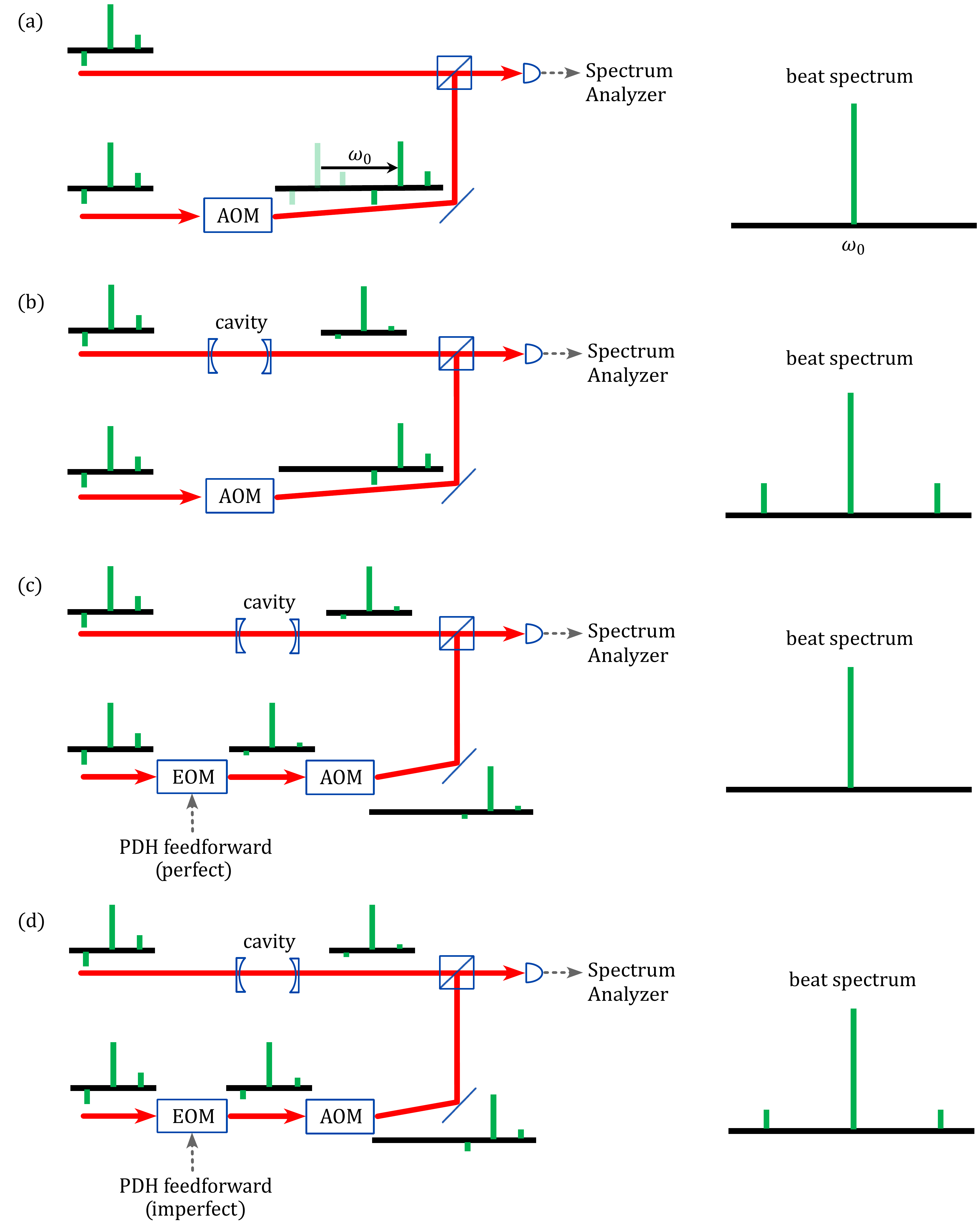}
	\caption{
		Illustration of heterodyne detection spectrum between the cavity filtered light and PDH feedforwarded light. (a) shows the spectrum when the two beating beams have identical noise sidebands but different carrier frequencies (one is shifted by $\omega_0$ using an AOM). In (b), one of the beams is filtered by an optical cavity. In (c) and (d), PDH feedforward is applied to the unfiltered beam. (c) assumes perfect PDH feedforward. This scenario appears to show perfect noise suppression since the noise suppression performance of PDH feedforward and cavity filter are identical. Imperfect PDH feedforward gives rise to the finite noise sidebands in (d).
	}\label{fig_heterodyne}
\end{figure}

In order to clarify the information contained in the spectrum of our heterodyne measurement, we shall consider four scenarios (Fig.~\ref{fig_heterodyne}). Fig.~\ref{fig_heterodyne} (a) shows the case when the two interfering beams have different carrier frequencies but identical phase-noise sidebands. In the insets, the two sidebands are drawn asymmetrically to emphasize that they have opposite phases. Using Eq.~(\ref{eq_Beat}), one can easily show that only the $\omega_0$ carrier can be observed in the beat spectrum, and the sidebands seem to disappear. When one of the beams is filtered using an optical cavity (Fig.~\ref{fig_heterodyne} (b)), the two interfering light fields have different sideband-to-carrier ratios. In this case, $\beta_{\rm trans}\sin(\omega_n t) \neq \beta_{\rm out}\sin(\omega_n t + \psi_{\rm diff})$ and thus $\phi_{\rm beat}^n \neq 0$, the noise sidebands become visible in the beat spectrum according to Eq.~(\ref{eq_Beat}).

In Fig.~\ref{fig_heterodyne} (c), PDH feedforward is applied to the unfiltered beam. In the ideal case, phase-noise suppression by PDH feedforward performs equally well as the cavity filter. Similar to the scenario in Fig.~\ref{fig_heterodyne} (a), the noise sidebands should also disappear from the beat spectrum, resulting in $A_{\rm heterodyne}=0$. However, due to imperfections in the experiment, the sidebands are visible in a real measurement (Fig.~\ref{fig_heterodyne} (d)), and this gives a finite $A_{\rm heterodyne}$.

As is explained above, what we can extract from the measured beat signal is $|\phi_{\rm beat}^n|$. Meanwhile, we can compute $|\phi_{\rm trans}^n|$ by using $|\phi_{\rm trans}^n| = \beta_{\rm trans} = \beta_n \sqrt{{\rm Re}\mathcal{T}(\omega_n)^2 + {\rm Im}\mathcal{T}(\omega_n)^2}$ (consistent with Eq.~(\ref{eq_Trans})). As the imperfections in applying feedforward can be caused by different reasons and vary for different frequency, we assume the phase $\theta$ can be of any values. Therefore, the maximum possible amplitude of $\phi_{\rm out}^n$ is $|\phi_{\rm trans}^n| + |\phi_{\rm beat}^n|$. Based on the above analysis, we compute the conservative lower bound of actual noise attenuation as
\begin{equation} \label{eq_Actual}
	A_{\rm actual,~lower bound}(\omega_n) = \left[\sqrt{A_{\rm heterodyne}(\omega_n)} + \sqrt{A_{\rm cavity \ filter}(\omega_n)}\ \right]^2.
\end{equation}
Here, $A_{\rm heterodyne}(\omega_n)$ is the noise attenuation parameter measured by self-heterodyne measurement, and $A_{\rm cavity \ filter}(\omega_n) = P_{\rm trans}(\omega_n)/P_{\rm in}(\omega_n)$. $P_{\rm in}(\omega_n)$ is the phase noise power incident to the cavity, $P_{\rm trans}(\omega_n)$ is the phase noise power transmitted through the cavity.

\section{Procedures for optimizing feedforward}

As is stated before, when one imprints a phase shift proportional to the residual PDH error signal $V_\mathrm{error}(t)\propto \sin\phi(t)$ to the light source, the output field becomes $E_{\rm out} \propto \exp\left\{i \left[ \omega_c t + \phi(t)+ G \sin\phi(t^{\prime})\right]\right\}$, where $t^\prime$ highlights the possible mismatch in the feedforward signal and the corresponding noise before EOM2. To achieve the best cancellation of phase noise, we need (1) to match $t$ and $t^\prime$ and (2) to set $G=-1$ for all frequencies. The procedures are as follows:
\begin{enumerate}
	\item Estimate the delay in the feedforward path $\tau$, approximately the time it takes for light to travel from the laser to APD1, and for the electrical signal to travel from APD1 to EOM2.
	\item Insert a delay fiber that would hold up the light by slightly more than $\tau$.
	\item Change the gain of the flat loop filter P to maximize suppression of the low-frequency phase noises shown in the self-heterodyne spectrum measured at SA (because feedforward suppression of low-frequency noise is more forgiving in the mismatch between $t$ and $t^\prime$).
	\item  Adjust the length of coaxial cable between the loop filter P and EOM2 to optimize the suppression performance at higher frequencies.
	\item Repeat step 3 and 4 to obtain best suppression for frequencies of interest.
\end{enumerate}

\section{Stability of PDH Feedforward}

\begin{figure}[!htbp]
	\centering
	\includegraphics[width=0.45\columnwidth]{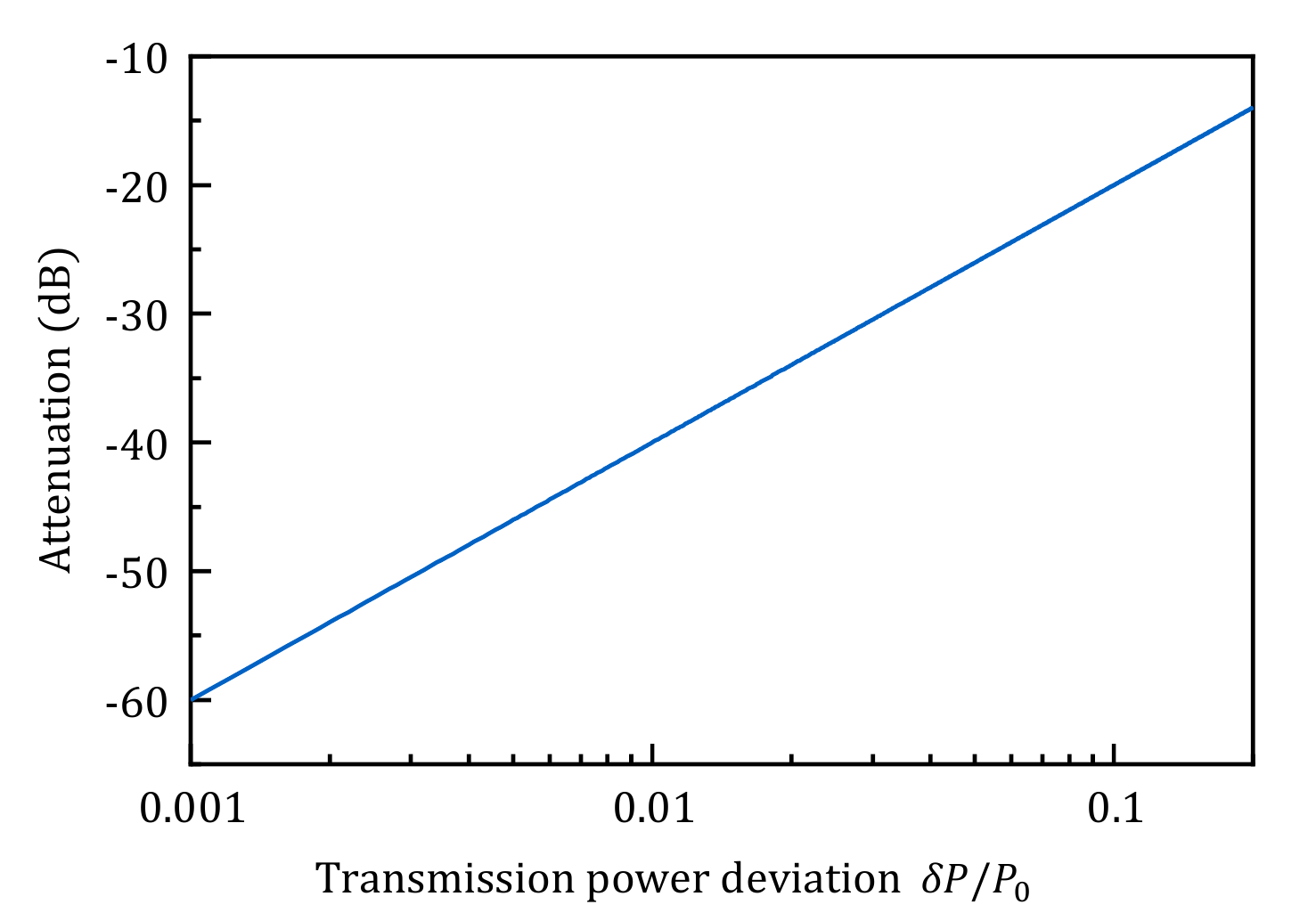}
	\caption{
		Phase noise attenuation as a function of deviation of the optical power transmitted through the cavity. $P_0$ is the power giving $G=-1$, and $\delta P$ is the absolute change of that.
	}\label{fig_stability}
\end{figure}

The performance of PDH feedforward depends on how well $t$ and $t^\prime$ match and how close the open loop gain $G$ is to $-1$. Because the delay time in the optical path and the electrical path for the feedforward signal is relatively stable, the long-term stability of PDH feedforward is limited mainly by the fluctuation of $G$. From Eq.~(\ref{eq_PDH_error}), we know that $G$ is proportional to $E_0^2 J_0(\beta) J_1(\beta)$, where $E_0^2$ is directly proportional to the power of light transmitted through the cavity. Therefore, $G$ would deviate from its ideal value when the power of the incident light drifts, and when the spatial mode matching between the incident light and the cavity deteriorates. Fig.~(\ref{fig_stability}) shows how the phase noise suppression performance worsens when the transmitted power deviates from the ideal value $P_0$. It indicates that even when the transmission varies by $10\%$, the noise attenuation would still be better than $-20$~dB. For better stability, we recommend monitoring the cavity transmission power and stabilizing it with a slow feedback loop. To maintain a phase noise attenuation better than $-40$~dB, the transmission drift should be smaller than 1\%.

\bibliographystyle{apsrev4-2}

\begin{thebibliography}{59}
\bibitem [{com()}]{comment_Freq_Ref}%
		\BibitemOpen
		\href@noop {} {}\bibinfo {note} {High-frequency noises above $\gamma_c$ are
			filtered out by the cavity, while low-frequency noises are strongly
			suppressed by PDH feedback, resulting in laser linewidth $\ll
			\gamma_c$.}\BibitemShut {Stop}%
\bibitem [{rea()}]{realcavity}%
		\BibitemOpen
		\href@noop {} {}\bibinfo {note} {For clarity of presentation, we assume a
			cavity with complete transmission at resonance in this manuscript. All
			presented results, figures and conclusions are nonetheless intact for real
			cavities with loss or imbalanced reflectivities.}\BibitemShut {Stop}%
\bibitem [{ref()}]{refractiveindex}%
		\BibitemOpen
		\href@noop {} {}\bibinfo {note} {When estimating the effective length, one
			should account for the refractive indices of all the relevant optical
			elements and coaxial cables, and also the delay caused by all the relevant
			electronic components.}\BibitemShut {Stop}%
\end{thebibliography}

%

\end{document}